\documentclass[12pt]{article}

\usepackage{pdfpages}
\pdfoutput=1

\def\cl{\centerline}

\def\sghr{{\hat \sigma}_{\lambda}^2}

\def\Th{\hat{T} }

\def\part{\partial } 
\def\qh{\hat{q}_{geo} }
\def\qhb{\overline{\hat{q}}}

\def\ybf{{\bf y}\,}
\def\zbf{{\bf z}\,}

\def\albf{{\bf \alpha}\, }
\def\alh{\hat{\bf \alpha}\, }

\def\mubf{{\bf \mu}\, }

\def\Sibf{\mathop{\mbox{\boldmath $\Sigma$}}}
\def\mubf{\mathop{\mbox{\boldmath $\mu$}}}

\def\Dbf{{\bf D}\,}

\def\Kbf{{\bf K}\,}

\def\nb{\overline{n} }
\def\psb{\bar{\psi} }

\def\hv{{\bf h}}
\def\uv{{\bf u}}

\def\np{\vfill\eject}       % new page
\def\ni{\noindent}
\def\IR{I\kern-.255em R}

\def\nb{\overline{n} }

\def\Xbf{{\bf X}\,}

\def\Sbf{{\bf S}\, }

\def\Cbf{{\bf C}\,}

\def\Gbf{{\bf G}_{\lambda}\,}

\def\Sbf{{\bf S}\,}

\def\albf{\mathop{\mbox{\boldmath $\alpha$}}}
\def\albfdag{\mathop{\mbox{{\boldmath $\alpha$}$^{\dag}$}}}
\def\epsbf{\mathop{\mbox{\boldmath $\epsilon$}}}
\def\albh{\hat{\mathop{\mbox{{\boldmath $\alpha$}}}}}

\def\mubf{\mathop{\mbox{\boldmath $\mu$}}}

\def\Sibf{\mathop{\mbox{\boldmath $\Sigma$}}}

\begin{document}
\begin{center}
{\bf A HIERARCHY OF EMPIRICAL MODELS OF \\
PLASMA PROFILES AND TRANSPORT}
\end{center}

\begin{center}
{\bf Kaya Imre and Kurt S. Riedel}\\
{$^{ }$ New York University, 251 Mercer St., New York  NY 10012-1185} \\
{\bf Beatrix Schunke} \\
{$^{ }$JET Joint Undertaking, Abingdon, Oxon, OX14 3EA, UK} \\
\end{center}

%\begin{abstract}
\centerline{\bf Abstract}

Two  families of statistical models of increasing statistical
complexity are presented which generalize global confinement expressions 
to plasma profiles and local transport coefficients. 
The temperature or
diffusivity is parameterized as a function of the normalized flux radius,
$\psb$, and the engineering variables, ${\bf u} = (I_p,B_t,\bar{n},q_{95})^{\dag}$. 
The log-additive temperature model assumes that
$\ln [T(\psb, {\bf u})] =$ $f_0 (\psb) + f_I (\psb)\ln[I_p]$ $+ f_B (\psb) \ln [B_t]$
$+ f_n (\psb) \ln [ \bar{n}] + f_{q}\ln[q_{95}]$. 
The unknown $f_i (\psb)$ are estimated using
smoothing splines. The Rice selection criterion is used to determine
which terms in the log-linear model to include. 
A 43 profile  Ohmic data set from the Joint European Torus 
[P.~H.~Rebut, et al.,  Nuclear Fusion {\bf  25}
1011, (1985)]
is analyzed and its shape dependencies are described.  
%A good measure of the performance of a particular model of tokamak transport
%is the  ratio of the root expected average square error (REASE) of the
%fit to that of fitting each temperature profile separately with 
%smoothing splines. 
The best fit has an average  error of  152 eV which is 10.5 \%
percent of the typical line average temperature. 
%16.8 \%, which 
The average error is less than 
the estimated measurement error bars.
%only 22 \% larger than
%fitting each profile separately.
The second class of models is log-additive
diffusivity models where $\ln [ \chi (\psb, {\bf u})] $ $=\ g_0 (\psb) + 
g_I (\psb) \ln[I_p]$ $+ g_B (\psb) \ln [B_t ]$ $+ g_n (\psb) \ln [ \bar{n} ]$.
 These log-additive diffusivity models are useful when the diffusivity is
varied smoothly with the plasma parameters. A penalized
nonlinear regression technique is recommended to estimate the $g_i (\psb)$.
The physics implications of the two classes of models,  additive
log-temperature models and additive log-diffusivity models, are
different. The additive log-diffusivity models adjust the temperature profile
shape as the radial distribution of sinks and sources. In contrast, the
additive log-temperature model predicts that the temperature profile 
depends only on the global parameters and not on the radial
heat deposition.

%We are unaware of any model of anomalous transport (such as Rebut's critical
%temperature gradient model) which is able to achieve comparable values
%of the REMSE. We strongly suspect that the EMRSE enhancement factor 
%for such ``first principles'' based transport models 
%is much larger than 50 \%. 
%We believe it is so large that no one in the predictive transport community 
%is willing to measure their fit errors in these terms.
%Our EMSRE inflation factor of 50 \% indicates the presence of model error 
%in the fit. Nevertheless, we consider this small enhancement in the REMSE 
%to be a major success given the diverse set of plasma conditions in the %
%database and the simplicity of our model.
%

%\end{abstract}

PACS NUMBERS: 02, 52.55Fa, 52.55Pi, 52.65+z

\ 

\

%\newpage
{\bf I. INTRODUCTION}

Global confinement expressions have proven useful in understanding and
predicting plasma performance$^{1-7}$. These confinement expressions are 
straightforward
to analyze statistically, but do not address the radial variation of the
plasma profiles and plasma transport coefficients. In this article, we
describe two families of empirical models which generalize global scaling
expressions to profiles and diffusivities.

We define the $m$ vector, $\uv$, to be a vector of global engineering
variables. Typically, the components of $\uv$ are the logarithms of the
edge safety factor, $q_{95}$, the plasma current, $I_p$ (in MA), the
toroidal magnetic field, $B_t$ (in Tesla),  
the line average density, $\bar{n}_e$ (in $10^{19}/m^3$), the
absorbed power, $P$ (in MW), the effective ion charge, $Z_{eff}$,
the isotope mixture, $M$, the plasma elongation,
$\kappa$, and the major and minor radii. Other engineering variables can
include the divertor configuration, the wall type, and the type of heating.
In practice, we usually work with the logarithms of the engineering variables,
and normalize the variables about their mean values in the data set.
In this notation, the standard power
law for the energy confinement time, $\tau_E$, is
$\tau_E = c_0 I_p^{\beta_1} B_t^{\beta_2}  \nb^{\beta_3} \ldots$, where
$\beta_{\ell}$ are the scaling exponents. The power law can be rewritten
as a log-linear expression
\begin{equation}\label{E1}
\ln \tau  = \beta_0 + \beta_I \ln[I_p]  + \beta_n \ln[\nb] + \ldots 
\ . 
\end{equation}  %\eqno (1)$$
In this form, the resulting scaling expression can be analyzed using
linear regression. Ordinary least squares analysis makes a number of
implicit assumptions which are described in Refs.~1 and 3.

We consider the plasma temperature as a function of the normalized
radial flux variable, $\psb$, and the plasma control variables, $\uv$.
A convenient flux variable normalization is that the toroidal flux through
a given radius, $\psb$, is equal to $\psb^2$ times the total flux. 
In this section, we neglect the
random errors associated with the data and concentrate on the empirical
model. By writing $T(\psb, \uv)$, we are implying that the temperature is
an unknown function of $m+1$ variables. Attempting to estimate an arbitrary
$m+1$ dimensional function from the measured tokamak data is a very
ill-conditioned problem. Therefore, {\it we restrict the class of models which
we examine to a more limited class.}

Experimentalists have often observed that the Ohmic temperature profile
shape varies very little as the plasma control variables vary. This ``profile
resilience'' motivates us to define ``profile resilient'' models:
\begin{equation}\label{E2}
\ln[T(\psb, \uv)] = f_0 (\psb) + H( \uv ) \ \ .  
\end{equation}                  %\eqno (2)$$
$H(\uv)$ is independent of  $\psb$ and  changes the magnitude but not
the shape of the temperature profile.
In the log-linear case, $H( \uv) = c_I \ln[I_p] + c_B \ln[B_t] + c_n 
\ln[\nb] + \ldots$.
We determine
$f_0 (\psb)$ and $H(\uv)$ by fitting $f_0 (\psb)$ with smoothing
splines and using linear regression. 

A more general model is to let the shape depend on $q_{95}$:
\begin{equation}\label{E3}
\ln[T(\psb, \uv)] = f_0 (\psb) + f_q (\psb) q_{95} + H( \uv) \ \ , 
\end{equation}              %\eqno (3)$$
where both unknown radial functions, $f_0 (\psb)$ and $f_q (\psb)$, 
are fit with smoothing splines. 
%Tang's transport model$^{10,11}$ is based on the 
%profile consistency model of Eq.~(3).  
Tang's well-known transport model$^{8,9}$, based on profile consistency, 
is a special case of model of Eq.~(\ref{E3}).  
Tang's model requires the log-temperature profile shape to be
be quadratic: $(f_0 (\psb) \equiv c_0 \psb^2 $ and  
$f_q (\psb) \equiv c_q\psb^2 )$,
and derives $H( \uv)$ from theoretical consideration. 

More generally, we have the additive spline model of Refs.~$10-12$:
\begin{equation}\label{E4}
\ln[T(\psb, \uv)] = f_0 (\psb) \ + H(\uv) + 
\sum_{\ell =1}^L f_{\ell} (\psb)h_{\ell}(\uv) 
 \ .   %\eqno (4)$$
\end{equation}
In Eq.~(\ref{E4}), we have separated  $f_0 (\psb)$ and  $H(\uv)$ from the other
terms to stress that these terms are the ``profile consistent'' terms.
If desired, additional cross-terms may be added to Eq.~(\ref{E4}).
 
In Section II, we describe our fitting procedure 
for estimating the free parameters in the log-additive
 model of  
Eq.~(\ref{E4}).
In Section III, we apply our method to  the Ohmic data 
from the  Joint European Torus$^{13}$ (JET) with
$\hv( \uv) = ( \ln[I_p], \ln[B_t], \ln[\bar{n}],$ $\ln[q_{95}])$.
The resulting model can be rewritten as
\begin{equation}\label{E5}
T(\psb) = \mu_0 (\psb) I_p^{f_1 (\psb)} B_t^{f_2 (\psb)}
\bar{n}^{f_3 (\psb)} q_{95}^{f_4 (\psb)} \ ,
\end{equation}                 %\eqno(5)$$
with $\mu_0(\psb) = \exp(f_0(\psb))$.
In place of  $q_{95}$ in Eq.~(\ref{E5}), we can use the geometric part of
the  safety  factor: % where $\qh$ is the geometric part of the safety factor:
$\qh \equiv q_{95}I_p/B_t$.  We also find that a simpler model with
``$f_1$, $f_2$, and $f_3 = $ constant'' fits the data to reasonable precision.
 
In Sections IV and V, we introduce a second family of models for the
log-diffusivity. We summarize our results in Sec.~VI. The
appendix  describes
our model selection criteria; i.e.~how we use the data to determine which
log-linear model is most appropriate. 
%Appendix B describes our estimate of the radial distribution of theestimation 
%local error.Appendix C describes our covariance model. %structure.

\ \\

\noindent
{\bf II. PROFILE ESTIMATION AND MODEL SELECTION}
%arameter Estimation Methods}

To estimate the unknown functions, $f_{\ell}(\psb)$ in the additive 
log-temperature models, we expand each of the functions in B-splines:
$f_{\ell}(\psb)=\sum_{k=1}^K \alpha_{\ell k}B_k(\psb)$, 
where the $B_k(\psb)$
are the cubic B-spline functions.  The $\alpha_{\ell k}$ are free parameters
which need to be estimated. 
In Refs. 10-12, we describe how estimation of the unknown functions in
the log-additive temperature model can be 
formulated as a large linear regression problem. In Ref.~12, we show how
a smoothness penalty function can be used to advantage in the spline
fit to the additive log-temperature model. 
We denote the fitted response function by
$\hat{T} (\psb , \uv | f_0 ,\ldots , f_L )$.
The algorithm of Ref.~12 is simply:
Minimize with respect
to the B-spline coefficients of $f_0 (\psb) , \ldots , f_L (\psb)$ of Eq.~(4),
the weighted least squares problem:
\begin{equation}\label{E6}
\sum_{i,j} \left| {T_i (\psb_j^i ) - \hat{T} (\psb_j^i , \uv_i | f_0 ,
\ldots , f_L ) \over \sigma_{i,j}}\right|^2 + \sum_{\ell = 0}^L \lambda_{\ell}
\int_0^1 |f_{\ell}^{\prime \prime \prime} (\psb) |^2 d\psb \ ,
\end{equation}         %\eqno (6)$$
where $T_i (\psb_j^i )$ is the $j$th radial measurement of the $i$th 
measured temperature profile and $\sigma_{i,j}$
is the associated error. The second term is the smoothness penalty which damps
artificial oscillations in the estimated $f_{\ell} (\psb)$.
The smoothing parameter, $\lambda_{\ell}$, controls the smoothness of the 
estimate of $f_{\ell}(\psb)$. The appendix describes how we  determine 
$\lambda_{\ell}$ empirically.

There are two types of systematic error: model error (since the additive
model is only an approximation), and smoothing error from the 
smoothness penalty function. Simplifying the physical model  causes
bias error, but can often reduce the variance of the fitted model.
This variance reduction occurs because the simplified model usually
has fewer free parameters than a more complete model does.

We wish to choose the additive model which minimizes the error in
predicting the temperature of a new profile. 
Unfortunately, the prediction error depends on the unknown ``true''
temperature function. To select the best model and smoothing parameters,
the expected average square error (EASE) is  estimated empirically$^{14,15}$. 
We use a generalization of the estimate of the EASE given by the
Rice criterion.~(See the appendix.) This EASE estimate includes
the bias error associated with the incomplete model, i.e.~we admit
that our additive model is systematically wrong, and we estimate the 
size of this error. 

From this estimate of the EASE, we then select
the additive model which minimizes the Rice criterion. %empirical EASE.
Similarly, we choose the smoothing parameters, $\lambda_{\ell}$, to minimize 
this empirical estimate of the expected error.   
%The smoothing parameters, $\lambda_{\ell}$ are chosen to minimize 
%an expected prediction error.  
The Rice criterion estimates the fit error for $new$ data while the older 
``$\chi^2$'' statistic considers the fit quality for the existing data set.
The Rice criterion is more selective than the $\chi^2$ statistic in the
sense that it prefers simpler, lower order models.

\ \\

\noindent
{\bf III. JET OHMIC TEMPERATURE PROFILE PARAMETERIZATION}
%Ohmic Temperature Profile Parameterization}

{\it a) Single Profile Analysis}

We consider a 43 profile data set  from  the Joint European Torus$^{13}$.
The electron temperature and density profiles are measured by the JET LIDAR
Thomson scattering diagnostic. Each profile is measured at approximately
50 radial locations along the plasma mid-plane.
Table 1 summarizes the global parameters of the data. 
The data contains discharges with the edge safety factor, $q_{95}$, as high
as 12.

The JET discharges were produced between 1989-90 and 1991-92. During this time,
JET %limiter 
operated with carbon tiles, with carbon tile and beryllium
evaporation, and with beryllium tiles. 
Most of the discharges in the database have the plasma boundary formed by the
outer wall limiter with beryllium tiles. 
%Figure 1 shows a typical plasma cross-section.

The JET LIDAR Thomson scattering diagnostic is described in Ref.~16.
A number of  the profiles have an artificial increase in the measured
temperature near the inner wall. This  problem occurs because the laser
light from the %Thomson scattering 
LIDAR diagnostic is partially reflected
near the plasma wall and 
stimulates radiation emission near the plasma edge. 
To prevent these spurious data from influencing 
the profile fit, we delete measurements in the outer ten percent of the
plasma which have the temperature increasing near the wall.
%In quoting our goodness of fit values, we use only measured temperatures
%above 350 eV.

Near the inboard wall, there are only rarely usable measurements. To be
able to estimate the temperature for $\psb < -0.87$, we reflect the temperature
for $\psb > +0.87$. In our model selection criterion, we use the number
of measured data points and not the number of  augmented data. 

Neither the fitted functions, $\hat{f_{\ell}}(\psb)$, nor the measured data
are symmetric with  respect to $\psb$. The measured profiles are clearly
hotter and broader at the inboard side (negative $\psb$). 
Since the  LIDAR measurements near the inner wall 
tend to be less accurate than those on the outboard side, we prefer to fit
the data with an asymmetric profile. If a symmetric fit is desired, we
recommend using our fit restricted to $\psb \ge 0$.
It is unclear if this asymmetry is due to a systematic error in the
flux map or in the LIDAR measurements or  has  an unknown physical cause.

Our fitted profiles allow asymmetry and reproduce this asymmetry.
We do test each $f_{\ell}$ separately for symmetry.  Making
%finding is that 
$f_1(\psb) \ldots f_4(\psb)$
symmetric in $\psb$ while keeping $f_0(\psb)$ asymmetric
in our best fit model raises the fit error only slightly from
150 to 152 eV.
If we force $f_0(\psb)$ to be symmetric, the difference in the residual
fit error is noticeable.

In our spline fits, we use 20 knots. To reduce the ill-conditioning
of the fit near the plasma edge, we decrease the density of knots 
near the edge.
Our profile fits depend only very weakly on the knot spacing due to
the smoothness penalty term. In contrast, if no smoothness penalty is
used (as in the original algorithm of Refs.~10-11), the fit is strongly
influenced by knot selection. Our data fit give an accurate fit to
the  data which is nearly independent of the knot positions.
%in the remainder of this section 
The smoothing spline yields accurate representation of the solution
with fewer artificial oscillations than the 
%demonstrate that our present fitting procedure constitutes a major
%improvement over 
methodology of Ref.~10.

%Figure 2 displays smoothing spline fits of two temperature profiles.
Fitting each profile separately gives a root mean square error (RMSE) of
.171   on the logarithmic scale, %the RMSE is ?, 
which corresponds to a relative fit error of 17.1\% .
On the linear scale the average fit error is 152 eV which is 10.5\%
%187 eV which is 12.8\%
percent of the typical line average temperature (1.454 KeV). 
The root mean square error (RMSE) is much larger (187 eV) than the
mean absolute error, indicating that a small percentage of the data points
are being fit very poorly. 
For the individual fits, the Rice criterion selects a relatively
small amount of smoothing and the fit tends to follow the 
small scale oscillations in the data.
%We are quoting
%our estimate of $\sigma_0$; the fit error at the  edge is somewhat larger
%because of our model of the measurement variance. (See Appendix B.)

\
 
{\it b) Model Selection}

We begin by considering the profile consistency model of Eq.~(\ref{E2}):
\begin{equation}\label{E7}
\ln [T] = f_0 (\psb) + c_I \ln [I_p ] + c_B \ln [B_t ]
+ c_n \ln [ \bar{n}] + c_{\kappa} \ln [ \kappa ] 
+ c_a \ln [a] \ .  %, \eqno (7)$$
\end{equation}
%where $c_1 = 0.7 \ldots$. 
This profile consistency model has a Rice
criterion value of 1.26. This is 42\% larger than our final
nonparametric fit in Eq.~(\ref{E8}). 
Thus, the spatial dependencies $\ln [I_p ]$,
$\ln [B_t ]$ and $\ln [ \bar{n} ]$ are significant.
We now consider nonparametric models which include spatial variation in
the control variable dependencies.

To select our log-linear model, we use a selection procedure based on the
Rice criterion. Our list of candidate variables is $\ln [I_p ]$,
$\ln [B_t ]$, $\ln [ \bar{n} ]$, $\ln [q_{95} ]$, $\ln[\kappa]$,
%$\ln[\qh \equiv q_{95}I_p/B_t]$,
$Z_{eff}$, $V_{loop}$, $a$, $R$, $\ell_i$, and time.
%and an indicator variable for beryllium versus carbon coating. 
At the $\ell$-th stages, we try all
possible combinations of the best model at the $( \ell -1)$-th stage
plus one additional variable. We choose the model which reduces the Rice
criterion the most.

Table 2 summarizes the reduction in the Rice criterion at each stage.
$\ln [I_p ]$ is clearly the most important control variable.
%followed by$\ln [B_t ]$ and then $\ln [ \bar{n} ]$. 
This result contrasts with
earlier profile consistency studies$^{12,17}$ which found that the
edge safety factor, $q_{95}$, is the most important variable in determining
the temperature profile shape. This difference occurs because we are fitting
the unnormalized temperature and $I_p$ is much more important than $q_{95}$
in determining the magnitude of the temperature. 

At the second stage of the sequential selection procedure,
the pair $(\ln [I_p ], \ln[B_t])$ minimizes the Rice criterion. 
%However, both $(\ln [I_p ], \ln[B_t])$ and $(\ln [I_p ], \ln[q_{95}])$
%achieve nearly the same value of the Rice criterion, and we consider these
%pairs further. At the third stage, we add each possible variable to each
%of the three candidate pairs. Of all the candidate models, 
In the second column of Table 2, we compare the two control variable model:
using $f_I (\psb) \ln [I_p ] + f_B(\psb) \ln[B_t ] $ 
with the two  control variable model and using 
$f_I (\psb) \ln [I_p ] + f_q(\psb) \ln[q_{95} ]$,
and we show that the Rice criterion is lower when $I_p$ and $B_t$ are used.
The same result holds when $\ln[\nb]$ is added to both models.
%We have tried replacing $f_1 (\psb) \ln [I_p ] + f_2 (\psb) \ln
%[B_t ] $ in Eq. (6) with $f_q (\psb) \ln [q_{95}] + c_B \ln [B_t ]$.
%The resulting Rice criterion was ?, showing that the temperature shape does
%not depend exclusively on the edge safety factor.
At the third stage,
$(\ln [I_p ], \ln[B_t],\ln[\bar{n}])$ has  the lowest value of the
Rice criterion.

At the fourth stage, $\ln[q_{95}]$ had the lowest value of the Rice criterion
when paired with the ``seed variables'',
$(\ln [I_p ], \ln[B_t],\ln[\bar{n}])$. Initially, we had  difficulty
accepting this result because the $I_p$ and $B_t$ dependencies of $q_{95}$
were already accounted for in the model and the geometric parameters,
$\kappa$, $a$ and $R$ vary very little. To test if this result were real,
we explicitly removed the  $I_p$ and $B_t$ dependencies from $q_{95}$
by defining $\qh \equiv q_{95}I_p/B_t$. 
%In an aspect ratio expansion,
%$\qh \simeq 5 a^2(1 +\kappa^2)(1.17 -.65a/R)/(2R)$ to good approximation.
We found that adding $\ln[\qh]$ resulted in an  even smaller value of the
Rice criterion than adding $\ln[q_{95}]$. This occurs because 
$f_{\qh}(\psb)$ varies less than $f_{q_{95}}(\psb)$. As a result,
we use fewer effective degrees of freedom to represent $f_{\qh}(\psb)$.

%As the fourth column of Table 2 indicates, $\kappa$ is the next most important
%variable. 
%Naive error bars do not include the uncertainty due to model misfit.
%(See Appendices A and B.)
%Since the Rice criterion measures model
%misfit as well, we have inflated
%the error bars by a factor of $\sqrt{C_R}$ to compensate for model error.

In general, adding a fifth variable resulted in little reduction in
the Rice criterion, and the new function, $f_5(\psb)$, would be nearly
constant with  large error bars. As a result, we stopped the log-linear
expansion with four  control variables.
Our final model is 
%Adding $\qh$, our best log-linear model is
$$
\ln [T] = f_0 (\psb) + f_I (\psb) \ln [I_p/I_0 ] + f_B (\psb) \ln [B_t/B_o ]
$$
\begin{equation}\label{E8}
+ f_n (\psb) \ln [ \bar{n}/n_o] + f_q(\psb)  \ln [\qh/\qhb ] 
%+ c_5  \ln [a] 
\ , %\eqno (7)
\end{equation}
where  $\qh \equiv q_{95}I_p/B_t$,
$I_o = 2.552$, $B_o = 2.710$, $n_o = 2.171$, and $\qhb = 4.150$.  
% qb= 4.537$.%$q_o = 4.537$.
Table 3 evaluates $f_0 (\psb) \ldots f_4 (\psb)$ at equispaced intervals. 
At the cost of increasing the fit error by 6 \% (from 150 to 159 eV),
we can replace the model of Eq.~(\ref{E8}) with the simpler  model 
$$
\ln [T] = f_0 (\psb) + f_q (\psb) \ln[q_{95}/4.537] +
c_I \ln [I_p/I_0 ] 
$$
\begin{equation}\label{E8Q}
+ c_B  \ln [B_t/B_o ]
+ c_n  \ln [ \bar{n}/n_o]
%+ c_5  \ln [a] 
\ , %\eqno (7)
\end{equation}
where $c_I$, $c_B$ and $c_n$ are independent of $\psb$. The best fit values
are $c_I = 0.69$, $c_B = 0.49$, $c_n = -.37$ and  $ f_q (\psb) \approx -0.2$
for $|\psb| < 0.66$ and decreasing to $-.37$ at the edge. 
This simpler model may be more robust than the best fit model of (\ref{E8}).
%Table 4 evaluates the coefficients in this model.
%$f_0 (\psb) \ldots f_4 (\psb)$ at equi-spaced intervals. 
The fit functions for both Eq.~(\ref{E8}) and  Eq.~(\ref{E8Q}) are available
from the authors.

%due to terms in the non-Spitzer 

{\it c) Fit Results}

Figure 1 plots  %the mean value of each of the control variables, 
$\exp(f_0 (\psb))$  and the
%while Figures 3b,c plot the control variable functions, 
$f_{\ell} (\psb)$. % through $f_4 (\psb)$. 
$\exp(f_0 (\psb))$ is the predicted temperature at $I_p=I_0$, $B_T=B_0$ etc.
%Appendix B discusses the error bar estimate.
%In Figure 3b,
{\it $f_I (\psb)$ shows that the temperature broadens and becomes somewhat
hollow with increasing current} while $f_B (\psb)$ shows the same effect with
decreasing toroidal magnetic fields. If $f_B (\psb) = c - f_I (\psb)$,
then the shape of the profile would depend only on the ratio,
$B_t/I_p$. %Figure 3b shows that 
Thus the shape depends primarily but
not exclusively on $B_t/I_p$. $f_B (\psb)$ is more peaked than 
$f_I (\psb)$ is hollow, which shows that a relative change in $B_t$
changes the shape more than the corresponding change in $I_p$.
The sequential selection procedure selected $I_p$ over $B_t$
in the first  step because $I_p$ varied more than $B_t$. Thus
using $I_p$ reduced the fit error more.
%Figure 3c shows that 
$f_{\qh} (\psb)$ is less peaked than either
$f_B (\psb)$ or $f_I (\psb)$, and therefore changing $q_{95}$ by changing
the geometry ($a$, $R$ and $\kappa$) only weakly changes the profile 
shape.

%Figure 3c shows that 
$f_n(\psb)$ and $f_{\qh}(\psb)$ are roughly constant, 
%within the error bars of our estimate i.e.
which means that $\nb$ and $\qh$ have  little effect %no significant influence
on the shape of the temperature profile. When
$f_n(\psb)$ and $f_{\qh}(\psb)$ are replaced by constants, the mean absolute
residual fit
error increases from 150 eV to 156 eV.  These constants are also plotted
in Figure 1.
%Therefore, we simplify the model of Eq.(7) by constraining the fit so that 
%$f_3(\psb)$ and $f_4(\psb)$ have no made radial dependence.
%%equal to constant values.
%With this constraint, the best fitting model is
%Since we plot $(f_{\ell} (\psb))$ 
%The model of Eq. (6) can be rewritten as
%$$T(\psb) = \mu_0 (\psb) I_p^{f_1 (\psb)} B_t^{f_2 (\psb)}
%\bar{n}^{c_3 } \kappa^{c_4}  \ .  %\eqno(8)$$
%The radial variation of $\mu_0 (\psb)$, $f_1 (\psb)$ and ${f_2 (\psb)}$
%are nearly identical to the corresponding curves in Fig.~3a and 3b.
%In Figure 3a,  $(f_1 (\psb))$ is displayed,
%and in Figure 3b,  $\exp(f_0 (\psb)) \ldots (f_3 (\psb))$ are displayed.
Figure 2a plots the fitted temperature  versus $\psb$ and $I_p$ at fixed
values of the other parameters. 
Figure 2b shows how  the fitted temperature  varies  with $B_t$. 

Our results differ from earlier ``profile consistency'' results because
we fit both the temperature shape and magnitude simultaneously;
i.e.~we do not normalize the data. 
The sequential selection procedure shows that the total plasma current 
% is the most important control variable in determining the temperature profile. 
is more important than the edge
safety factor in determining the JET Ohmic temperature profile.  
%For the Ohmic fit,
%$f_I (\psb)$ shows that the temperature broadens and becomes somewhat
%hollow with increasing current. $f_B (\psb)$ shows the same effect with
%decreasing toroidal magnetic fields. 
%Since $f_{\kappa}(\psb)$ is  nearly constant, increasing plasma elongation 
%improves the average temperature, but does not change the temperature shape.
From the shape of the
$f_{\ell} (\psb)$, we see that the polynomial models of the radial
dependence poorly approximate the actual shape. %The shape functions
%$f_{\ell} (\psb)$ tend to be roughly constant in the interior 
%and then vary more strongly near the plasma edge. 

Figure 3 plots two of the fitted profiles to illustrate the goodness
of fit. Our fitted curve is generally inside the experimental error bars. 
The combined fit to Eq.~(\ref{E8}) gives a  mean absolute error %(RMSE) 
of 152 eV. Since the mean line averaged temperature is 1.454 KeV,
this is a 10.5 \% typical error. %for $T > 350 eV$. 
On the logarithmic scale, the RMSE is .171, which
corresponds to a relative fit error of 17.1 \%. 

The mean square error is relevant when the errors have a Gaussian 
distribution. In our fit, a small percentage of the data has much larger
residual errors than is typical. Averaging the square error instead of
averaging the absolute error inflates the influence of the poorly fitting
points. We believe that the mean absolute residual is a more relevant 
description of the quality of fit.

The Rice criterion value of 0.88 usually
means that the expected square error 
in predicting new data is 0.88 times larger than the  experimental 
variance. For independent errors, the Rice value should be
greater than one. Due to the oversampling of the LIDAR diagnostic,
 the measurement errors are autocorrelated, and we are able to fit
the data with smaller residual  fit error. In the appendix, we derive
a correction for the autocorrelation. Nevertheless, we ascribe the
smallness of our Rice value to the spatial autocorrelation and
possibly to uncertainties in the experimental error bars.
%more accurate predictions of the measurement
%Since $\sqrt{1.5} \simeq 1.22$,
%our prediction is on average only 22 \% worse than the  measurement error.
The Rice value of 0.88 is surprisingly small,  
given the simplicity of our model and the
diverse set of plasma conditions in the database.
Thus, we consider this small enhancement to be a major success. 

%{\it The expected root mean square error (REASE) of this fit 
%is only 50 \% larger than fitting each profile separately!}
%The factor of 50 \% indicates the presence of model error in the fit.
%Nevertheless, we consider this small enhancement in the REASE to be 
%a major success given the diverse set of plasma conditions in the database
%and the simplicity of our model. To our knowledge, {\it no model of
%tokamak transport comes close achieving  this small value of REASE over
%an entire range of plasma parameters.}  

A surprising result of our analysis is that {\it $\ell_i$ is not particularly
useful in estimating the temperature}. $\ell_i$ is the measured
value of the second moment of the poloidal magnetic field.
If we assume that the current distribution is given by Spitzer resistivity
(with a constant, spatially uniform $Z_{eff}$ profile)
then $\ell_i$ can be related to a spatially weighted moment 
of the temperature distribution. 
%which corresponds to a moment of the current profile when 
Thus, we would expect that larger values of $\ell_i$, corresponding to
peaked current profiles, would correlate with peaked temperature profiles. 
Our empirical observation of only a 
weak dependence of the temperature shape on $\ell_i$ shows that the
current and temperature profile shapes are partially decoupled. 
This could be due to variation in $Z_{eff}$ or due to  the  empirical
resistivity differing from the Spitzer value.

At the fifth stage,   ``time'', as measured from the beginning of the
discharge, is the next most important variable. %Since ``time'' is a 
%nonphysical variable, this result is somewhat surprising. 
We would like to restrict our analysis to time points in the flat-top.
However, there are time points in the early phase of the  current ramp down.
Adding  a time variable to our regression analysis corresponds to the
$\ln [T](\psb,t) \sim f_0 (\psb) + f_t (\psb)(t - \bar{t})$. 
 Our estimate of $f_t (\psb)$ shows that
earlier times tend to be more peaked and later times are flatter.
More specifically,  %The resulting profile function, 
$f_{time}(\psb)$ strongly resembles $f_{I}(\psb)$, which
indicates that  the profile shape is influenced by the time history
of the plasma current.
Although our log-additive
model is designed for the steady state part of the discharge,
most of these discharges have a similar time history.
%we have approximated the time dependence with the  additive ansatz: 
%$\ln [T](\psb,t) \sim f_0 (\psb) + f_t (\psb)(t - \bar{t})$ is a very crude
%ansatz. 
This explains why a nonphysical variable like ``time'' could reduce the fit
error.
One of the principal disadvantages of our log-linear temperature
models is that the profile shape does not adjust in a physical manner
when different time evolution scenarios are used. Thus, $f_t (\psb)$ is
an artifact of the standard time history scenario in JET, 
%Nevertheless, 
%$f_t (\psb)$ does reflect the typical temporal evolution of a JET temperature 
%profile.
%Since ``time'' is not a physical variable, 
and we reject using time as a control variable.
%we do not add ``time'' to the log-linear model. 

The electron temperature tends
to be hotter with carbon tiles than with beryllium tiles on the limiter. 
Our empirical
fit given in Eq.~(\ref{E8}) fits both classes of discharges, 
but the fit parameters
were basically determined by the 37 beryllium discharges. Given a larger
data set, we could quantify the systematic differences between 
carbon and beryllium.
%Thus we conclude
%that the differences in temperature are due to differences in the control variables.
There are three divertor discharges in our database. The other 40
profiles are limited by the
outer wall. Thus the free functions in our empirical
fit of Eq.~(\ref{E8}) are determined primarily by the limiter discharges.
%The value of the Rice criterion measures the multiplicative factor for the
%residual variance of the log-linear model relative to the residual fit variance
%from fitting each profile separately.
We caution that our results are based on a limited database of 43 JET
profiles and that other subsets of the JET data could show different
dependencies. 

\vspace{.5in}

\noindent
{\bf IV. SEMIPARAMETRIC MODELS OF THE DIFFUSIVITY}

Much effort has been devoted to determining the anomalous heat diffusivity
as a function of the local variables. Hundreds of anomalous transport
models have been proposed, none of which is widely accepted. In contrast,
there is a consensus of what the ``stereotypical'' heat diffusity looks like:
the heat diffusivity is usually flat in the inner half of the plasma
radius and then increases parabolically in the outer half of the plasma.
Furthermore, the radial variation of the heat  diffusivity 
profile appears to depend only weakly 
on the plasma parameters.
%While attending numerous  seminars and talks on plasma transport, 
%we have been impressed by how 
There are  exceptions to this general assertion, but we believe that
broad characterization of the heat diffusivity profile has been supported by
many empirical studies.
%is basically correct. More generally,

In constructing empirical models of the anomalous heat diffusivity,
our basic hypothesis
is that the single most important variable for plasma transport analysis 
is the normalized flux radius. In other words, normalized flux radius
is a more important control variable than more physical variables such as
the poloidal gyro-radius. % and the critical temperature gradient.
This assertion is difficult to prove or disprove, so we content ourselves
with describing a family of models which are based on this hypothesis. 
We also wish to parameterize the observed heat diffusivity as a function
of the engineering variables in order to influence design studies.

Thus, we propose a second class of models which 
is similar to the temperature models except
that we model the log-diffusivity. 
A ``diffusivity consistent'' model is
\begin{equation}\label{E9}
\ln [\chi({\psb},\uv)] = g_0 (\psb) + H( \uv) \ .
\end{equation}                 %\eqno (8)$$
Equation (\ref{E9}) implies that as a result the shape of $\chi$ depends solely
on radius and  that the shape of diffusivity is independent of both the local
and global plasma parameters including the temperature gradient. 
Thus, the model is Bohm-like for radial variation.
We believe that models
similar to Eq.~(\ref{E9}) approximate the experimental data fairly well
in the sense that the diffusivity tends to be flat out to $\psb = .6$, and
then increases parabolically. {\em We have not yet modeled plasma discharges
by parameterizing the diffusivity, but similar models have been used in plasma
modeling$^{19}$.}

In general, the radial variation of $\ln [ \chi ]$ can be represented as a
slowly varying function of flux radius and the engineering variables.
%depend on $q_{95}$ and the other control variables. 
Therefore, we generalize Eq.~(\ref{E9}) to all
log-additive models of diffusivity:
\begin{equation}\label{E10}
\ln [ \chi (\psb, \uv)] = g_0 (\psb) + \sum_{\ell =1}^L
g_{\ell} (\psb) h_{\ell} ( \uv) \ .          %\eqno (9)$$
\end{equation}
The inward density pinch can be included with a similar linear model.
As in Eq.~(\ref{E4}), we assume that the $h_{\ell} ( \uv )$ are known and are
typically $\ln [I_p] \ , \ln [\nb]$ and $\ln [B_t]$. 
The $g_{\ell} (\psb)$ are usually given by
other smoothing splines or low order polynomials.

The physics implications of the two classes of models -- additive
log-temperature models and additive log-diffusivity models -- are
different. The additive log-diffusivity models adjust the temperature profile
shape as the radial distribution of sinks and sources. In contrast, the
additive log-diffusivity model predicts that the temperature profile shape
does not depend only on the global parameters and not on the radial
distribution of sinks and sources. 

Profile resilience can be interpreted as the observation that the truth
is somewhere between these two viewpoints. In other words, the temperature
profile shape adjusts less than one would expect from a diffusive model$^{18}$.
In future work, we hope to compare the two classes of models to see
whether additive shape models better describe the temperature on the
diffusivity.

%A special case of the additive log-$\chi$ model is
%$$\ln [ \chi (\psb ,\uv)] = f_0 (\psb) + f_q (\psb)  + g( \uv) \eqno (7)$$
%In this case, the radial variation of $\chi$ depends only on the edge $q$.
%Tang's theoretical based transport model yields a diffusivity of the form
%given by Eq.~(7). Tang is able to simultaneously satisfy Eqs.~(3) and (7)
%because he considers a special theoretical input power deposition profile.

We can add additional variables to the control variable vector, $\uv$, such as
the beam penetration depth normalized to the minor radius, which partially
specify the heating profile. In this way, we can have additive
log-temperature models adjust to heating profiles and additive
log-diffusivity models be more profile resilient.

In the next section, we discuss the underlying difficulties in parameterizing
the diffusivity.

\vspace{.5in}

\noindent
{\bf V. ESTIMATION OF THE ADDITIVE LOG-DIFFUSIVITY MODEL}

\medskip

To estimate the B--spline coefficients for the additive log-diffusivity
model, we minimize Eq.~(\ref{E5}) as well. 
For the additive log-temperature model,
the predicted values, $\Th(\psb_j^i ,\uv_i |g_{\ell})$, are a linear function
of the spline coefficients and the resulting functional is quadratic. In
contrast, in the additive log-diffusivity model, 
$\hat{T} (\psb, \uv|g_{\ell} )$
is a nonlinear function of the unknown spline coefficient and each
evaluation of $\hat{T} (\psb,u|g_{\ell})$ in the minimization of Eq.~(\ref{E6})
requires the solution of the transport equation using a code such as 
SNAP$^{20}$. % or TRANSP.
The sinks and sources may be calculated for each discharge separately prior
to  beginning the least squares fit for the  additive log-diffusivity model.
Thus the evaluation of $\hat{T} (\psb,u|g_{\ell})$ requires only 
the inversion of a
heat transport equation with {\it fixed sinks and sources} at each step of
the minimization of Eq.~(\ref{E6}).

We are unaware of any  local $\chi$/heat flux
regression study that has  attempted to use $\psb$ and the global engineering
variables with a model similar to Eq.~(\ref{E10}).
Instead, researchers have tried to determine the dependencies of the
diffusivity on heat flux by regressing the point estimates of $\chi$ or
$\chi \nabla T$ versus local quantities. Previous local $\chi$/heat
flux regressions have ignored $\psb$ and have concentrated on 
local quantities such as the poloidal gyro-radius.
%critical temperature gradients. 
This approach has several
disadvantages relative to the additive log-diffusivity approach with a
global minimization. First, { we believe that the most useful variable in
fitting the diffusivity shapes is the normalized plasma radius,
$\psb$.}   Second, we believe that usually 
$\ln[\chi]$ has a simple and smoothly varying radial dependence. Third, 
the errors in estimates of 
the plasma gradients are often comparable to the errors in $\chi$ and
larger than the errors in the heat flux. When the dependent variables
have errors, linear regression is an inconsistent estimator of the
parameters. Even worse, the errors of the dependent and independent
variables are strongly correlated. Our personal experience is that
when $\chi$ is regressed  against $\nabla T$, the most likely result is
that $\chi \ \sim \ {I_pV_{loop}  \over n \nabla T}$. 
This result is strikingly similar to the definition of $\chi$. 
Note the similarity of this expression and 
the Coppi-Gruber-Mazzucato formula$^{21}$.

In the last paragraph, we described the problems in regressing the local
heat flux/$\chi$ versus the local plasma parameters instead of radius %$\psb$
and the global plasma parameters. We now describe a second set of
disadvantages which persist even when the independent variables are $\psb$
and $\uv$. %One of these advantages is that 
First, by fitting with Eq.~(\ref{E9}),
we force $\chi (\psb, \uv)$ to be a smoothly varying function. %In contrast
If the estimated $\chi$ is regressed at each radial point separately,
$\hat{\chi}$ will usually have spurious spatial oscillations.
Another severe disadvantage of regressing the local heat flux/$\chi$
directly is that the heat flux is measured and not inferred. Changing
$\hat{\chi} (\psb  , \uv)$ at one spatial location will modify the
predicted temperature, $\hat{T}$, at all radial locations. Thus to attain a
self-consistent estimation of $\chi$, we are forced to fit all radial
locations simultaneously. Finally, estimates of the variance of the
point estimate of $\hat{\chi}_i (\psb_j^i )$ are difficult to obtain, 
and this makes
a pointwise weighted least squares analysis usually infeasible.

For all these reasons, we prefer the global minimization approach of Eq.
(\ref{E5}) with the simple additive model of Eq.~(\ref{E10}). 
Nevertheless, the computational and programming effort to fit 
the log-additive $\chi$ model of Eq.~(\ref{E10}) is considerable and 
we have not yet applied it to JET data. 

\vspace{.5in}
%\np

\noindent
{\bf VI. SUMMARY}

Profile resilience and diffusivity profile resilience strongly
suggest that the appropriate empirical models for local profile
dependencies are the additive log-temperature model and additive
log-diffusivity model. We therefore distinguish four  classes of empirical
transport models:

\ni
1) {\it Global confinement models:} $\tau_E({\rm engineering\ variables})$ 
as typified by Eq.~(1);

\ni
2) {\it Semiparametric profile models:} $T(\psb,{\rm engineering\ variables})$
 as typified by Eq.~(\ref{E8});

\ni
3) {\it Semiparametric diffusivity models:} 
$\chi(\psb,{\rm engineering\ variables})$ 
as typified by Eq.~(\ref{E10});

\ni
4) {\it First principles transport models:} $T({\rm physics\ variables})$, 
possibly given by a theoretical expression.

The huge multiplicity of transport theories and the relative lack of
success in applying theory based models motivates us to consider the
semiparametric models of classes 2 and 3. 
We hope to use these same models to predict the profile peakedness factor for
deuterium-tritium
(D-T) discharges in JET and the tokamak fusion test reactor$^{22}$
(TFTR) and for extrapolating performance to International Tokamak Experimental
Reactor$^{23}$ (ITER).

We have accurately parameterized the JET Ohmic temperature
profiles using a log-linear temperature model (Class 2).
We have not yet fitted log-additive diffusivity models 
to the JET data (Class 3). In both cases,
the smoothing spline coefficients are best determined
by minimizing the residual fit error over all measured profiles
simultaneously.

%The additive log-temperature model has been
%successful in parameterizing the Ohmic profiles in ASDEX, JET and TFTR. 

%The primary result of our fit procedure is Table 3 which evaluates the
%profile parameterization at equi-spaced intervals. This 
Our parameterized temperature model, Eq.~(\ref{E8}), fits
our JET data set with a mean absolute error of 47 eV which is 3.2 \%
%RMSE error of 187 eV which is 12.8 \%
percent of the typical line average temperature.
%a relative accuracy of \%. 
We recommend using this
parameterization for transport analyses and most other applications.
%We are cautious as to whether our profile parameter should be used in 
%magnetohydrodynamic (MHD) stability studies because MHD stability is 
%sometimes sensitive to minor changes in the profile shape.

Is the database big enough to make  these conclusions?
Some of our conclusions will depend on the choice of data. If the 
database were to contain only discharges from a specialized scan
on one or two consecutive days, we would probably be able to fit 
the data better, including detailed profile features. Because
our database is taken over a diverse set of discharge conditions,
we average over the small scale features which depend on the particular
discharge conditions.  As a result, the fit is worse, but the results
are more robust because they reflect many different types of discharges.

%A good measure of the performance of a particular model of tokamak transport
%is the  ratio of the  root expected average square error (REASE) of the
%fit to that of fitting each temperature profile separately with 
%smoothing splines. Our best fit has an REASE which is 
%only 50 \% larger than fitting each profile separately!
%We are unaware of any ``first principle'' model of anomalous transport 
%%(such as Rebut's critical temperature gradient model) 
%which is able to achieve comparable values 
%of the REASE. We strongly suspect that the REASE enhancement factor 
%for such ``first principles'' based transport models 
%is much larger than 50 \%. 
%%We believe it is so large that no one in the predictive transport community 
%%is willing to measure their fit errors in these terms.

The Rice criterion measures the expected error in predicting new data,
normalized to the variance of the measurements for independent errors. 
Our Rice value of 0.88
means that our predictions are theoretically more accurate than
the experimental measurements.
%on average only 22 \% worse than the  measurement
%uncertainty.       %Our REASE inflation 
%The factor of 22 \% indicates the presence of model error in the fit.
This small value of the Rice criterion is probably due to
the spatial autocorrelation from oversampling and
possibly to uncertainties in the experimental error bars.
We consider this small value  of the predictive error 
%enhancement %in the REASE 
to be a major success given the diverse set of plasma conditions 
in the database and the simplicity of our model.

Our sequential selection procedure shows that the current is the most
important control variable in determining the temperature profile. Previous studies,
which have considered the normalized temperature, have found that the edge
safety factor is the most important control variable. Increasing the plasma current
results in broader, often hollow profiles. Fig.~1 shows that the current 
and the magnetic field modify the profile shape in different ways. Thus,
the temperature shape does not depend exclusively on  $q_{95}$.

%From the shape of the
%$f_{\ell} (\psb)$, we see that the polynomial models of the radial
%dependence poorly approximate the actual shape. The shape functions
%$f_{\ell} (\psb)$ tend to be roughly constant in the interior 
%and then vary more strongly near the plasma edge.

The profile parameterization in Eq.~(\ref{E8}) is only 
for time-independent profiles.
The log-linear diffusivity model may be more relevant to modeling time 
evolution. 
Since our database consist almost exclusively of limiter
discharges, we are unable to determine if the temperature profile is
modified when a divertor is used.

\ \\

{\bf APPENDIX: RISK ESTIMATION AND MODEL SELECTION}

The estimation of risk/expected error is critical to our analysis 
because we use this estimate to
select which terms to include in our analysis. We represent the ``true''
log-temperature values by the vector $\mubf$ and 
the measured log-temperature by ${\bf y} \equiv \mubf + \epsbf$.
We present the generalized cross-validation (GCV) estimate
of the  expected average square
error (EASE) as well as the Rice criterion correction. 
We consider the linear regression model:
%\begin{equation}\label{A1}
$$
{\bf y} =  \mubf + \epsbf \ ,
\ \ \ \ \mubf  := {\bf X} \albf
\ ,\eqno (A1)
$$   %\end{equation}    
where ${\bf y}$ is the measurement vector,
$\bf X$ is the data matrix, $\albf$ is
the parameter vector and $\epsbf$ is a vector of random
errors with covariance matrix $\Sibf$.
We define $\Dbf$ to be diagonal matrix which
contains the inverses of the variances of the measurements: 
$\Dbf_{i,j}= \Sigma_{i,i}^{-1} \delta_{i,j}$. 
Presently, we do not include the
off-diagonal terms in $\Sibf$ in the minimization, but do compensate
for this in our model selection criterion.

By $\mubf : = {\bf X} \albf$, we mean that we model $\mubf$ by ${\bf X}
\albf$, but that we admit that $\mubf = {\bf X} \albf$ is not exact and
that this model has a systematic error.
We estimate $\albf$ using the penalized
least squares estimate:
$$ %\begin{equation}\label{A2}
\hat{\albf}_{\lambda} = \arg\min_{\alpha}\left\{ 
({\bf y}-{\bf X}\albf)^{\dag}\Dbf ({\bf y}-{\bf X}\albf)    
%{({\bf y}-{\bf X}\alpha )^{\dag}({\bf y-X}{\alpha} ) \over \sigma^2} 
+ \lambda \albfdag {\bf S} \albf \right\} \ ,\eqno (A2)
$$ 
where $\bf S$ is the penalty matrix. 
Equation (A2) is an abstract matrix formulation of Eq.~(\ref{E6}).
For brevity, we denote $\zbf^{\dag}\Dbf \zbf$ by $||\zbf||^2_{\Dbf}$
where $\zbf$ is an arbitrary $n$-vector.
 
The subscript $\lambda$ on $\hat{\albf}_{\lambda}$
denotes the dependence on the smoothing parameter. Equation (A2) can be
rewritten as
$$ %\begin{equation}\label{A3}
\hat{\albf}_{\lambda} = [{\bf X}^{\dag}\Dbf {\bf X} +  \lambda {\bf S}]^{-1}
{\bf X}^{\dag} {\bf Dy} \ =\ \Gbf
{\bf X}^{\dag} {\bf Dy} \
,\eqno (A3)
$$
where ${\Gbf} \equiv [ \Xbf^{\dag}\Dbf {\Xbf }\
+  \lambda {\bf S}]^{-1}$.
The covariance of $\hat{\albf}_{\lambda}$ is
%the estimated spline coefficients is
$$     %\begin{equation}\label{A4}
{\bf Cov} [\hat{\albf}_{\lambda}\hat{\albf}_{\lambda}^{\dag}] =
\Gbf \Kbf \Gbf \ ,
\eqno(A4)$$
{\rm where} ${\bf K} \equiv {\bf X}^{\dag} {\bf D\Sibf DX}$.
In addition to the variance, Eq.~(A3) has a  bias/systematic error:
${\bf E}[{\mubf}-{\Xbf}\hat{\albf}_{\lambda} ]$.
%s the bias/systematic error.
The dominant source of bias error in our analysis
is due to model error in the additive model.
The expected average square error (EASE) in the fit
is
$$         %\begin{equation}\label{A5}
EASE=
{\bf E}[|| \mubf -{\bf X}{\albh_{\lambda} } ||^2_{\Dbf} ] 
%{\bf E}\left[\left({\bf \mu-X} \hat{\alpha}_{\lambda}\right)^{\dag} \Dbf^{-1}
%\left({\bf \mu-X} \hat{\alpha}_{\lambda}\right)\right]
= {\rm Bias}^2 + {\rm Variance} =
\left||{\bf E}[ \mubf - {\bf X}\hat{\albf}_{\lambda} ]\right||_{\Dbf}^2
+ {\rm trace} [\Cbf \Gbf_{} \Kbf \Gbf_{}]
\ ,\eqno (A5)
$$
where ${\bf C} \equiv {\bf X}^{\dag} {\bf DX}$ and
$ {\rm trace} [\Cbf \Gbf_{} \Kbf \Gbf_{}] =
{\rm trace} [\Dbf\Xbf {\bf Cov} 
[\hat{\albf}_{\lambda}\hat{\albf}_{\lambda}^{\dag}] \Xbf^{\dag}]$.
%Equation (A5) is derived by substituting Eq.~(A4) into Eq.~(A3) and computing

We wish to minimize the EASE. However, it is unknown and needs to be estimated.
We denote the average square residual of the empirical fit by
$\sghr(\lambda) \equiv 
{\left||{\bf y}-{\bf X} \hat{\albf}_{\lambda} \right||^2
/N}$.
The expectation of the square residual error is
$$      %\begin{equation}\label{A6}
{\bf E}\left[\left||{\bf y-X} \hat{\albf}_{\lambda}\right||^2_{\Dbf}\right]
%{\bf E}\left[\left({\bf y-X} \hat{\alpha}_{\lambda}\right)^{\dag} \Dbf^{-1}
%\left({\bf y-X} \hat{\alpha}_{\lambda}\right)\right]
= {\rm Bias}^2
+ \{  {\rm trace}[\Sibf \Dbf]\ -\ 2 \ {\rm trace}[\Kbf \Gbf]
+ {\rm trace} [{\Cbf \Gbf\Kbf\Gbf}] \}  \ .
\eqno (A6)$$
Note ${\rm trace}[\Sibf \Dbf]= N$.
Equation (A5) computes the error relative to the true, unmeasured values
while Eq.~(A6) uses the measured residuals. As a result,
Equation (A5) can be easily estimated from the data by computing the MSE
of the fit using the
measured temperature. In contrast, Eq.~(A5) involves the unknown, ``true''
temperature.
The Craven-Wahba estimate of the EASE uses Eq.~(A6) to estimate Eq.~(A5):
%to is simply
$$      %\begin{equation}\label{A7}
\widehat{{\bf E}[|| \mu -{\bf X}\hat{\albf}_{\lambda}||^2_{\Dbf} ]}
= ||{\bf y}-{\bf X} \hat{\albf}_{\lambda} ||^2_{\Dbf} - 
\{ N-2 \ {\rm trace}[\Kbf \Gbf] \}  \ .\eqno (A7)
$$
Equation (A7) is particularly valuable because it includes the systematic
error in directions which are orthogonal to the column space of $\bf X$; i.e.
the bias from not including all possible terms in the additive model.
At $\lambda=0$, ${\rm trace}[\Cbf \Gbf]$ equals the number of
fit parameters and decreases monotonically with $\lambda$.   %\over N} \ \ , \
Using $\Kbf$ in place of $\Cbf$ compensates for the autocorrelation of
the measurements. %When the distance between measurements is much less
%than the knot distance, ${\rm trace}[\Kbf \Gbf]\  \sim \ 
%\left(\sum${\rm trace}[\Cbf \Gbf]$$

The minimum of Eq.~(A7) with respect to $\lambda$ satisfies:
$$    % \begin{equation}\label{A8}
\part_{\lambda} \sghr + \frac{2}{N} \part_{\lambda}
{\rm trace}[\Kbf \Gbf]
=\ 0  \ .\eqno (A8)$$
Equation (A8) is useful when $\sigma^2$ is known. We now consider the case
where the covariance of the measurements is known up to an arbitrary
constant:  $Cov[\ybf \ybf^{\dag}]\ =\ \sigma^2 \Sibf$, where
$\sigma^2$ is unknown and $\Sibf$ is known. We continue to 
define $\Dbf_{i,j}= \Sigma_{i,i}^{-1} \delta_{i,j}$ and keep the  same
definitions for $\Gbf$, $\Cbf$ and $\Kbf$.  
When $\sigma^2$ is unknown, we can estimate it using
the Craven-Wahba estimate of $\sigma^2$:
$$     %\begin{equation}\label{A9}
\widehat{\sigma^2_{CW} } =
{\left||{\bf y-X} \hat{\albf}_{\lambda}\right||_{\Dbf}^2 %\right]
\over \left( N- \ {\rm trace}[\Kbf \Gbf] \right) } 
\ .\eqno(A9)$$
In penalized regression,  ``$N- \ {\rm trace}[\Kbf \Gbf]$'' 
is referred to as the effective number of degrees of freedom.

The empirical estimate of Eq.~(A8) using the estimate 
$\widehat{\sigma_{CW}^2}$ is
$$          %\begin{equation}\label{A10}
\part_{\lambda} \sghr + \frac{2 \sghr  }{N- {\rm trace}[\Kbf \Gbf]} 
\part_{\lambda} {\rm trace}[\Kbf \Gbf] 
%\part_{\lambda} \left\{ \frac{N\sghr}{
%\left( N- {\rm trace}[\Cbf \Gbf]\right)^2 } \right\}
=  0  
\ ,\eqno (A10)$$
which implies 
$$ %\begin{equation}\label{A11}
\part_{\lambda} \left\{ \frac{\sghr}{\left( 1- {\rm trace}[\Kbf \Gbf]/N
\right)^2 } \right\}
=  0  
\ .\eqno (A11)$$
Thus, we define the generalized cross-validation statistic as
$$ %\begin{equation}\label{A12}
GCV \equiv N
 \frac{||{\bf y}-{\bf X} \hat{\albf}_{\lambda} ||^2_{\Dbf}}
 {(N-  {\rm trace}[\Kbf \Gbf])^2 }  \ .\eqno (A12)
$$
Minimizing the GCV criterion of Eq.~(A12)
%Eq.~(A10) is an unbiased estimate of the expected error, it 
sometimes undersmoothes and tends to pick models with too many 
free parameters$^{15}$. 
We refer the reader to Ref.~15 for an empirical comparison of the 
Rice criterion and the GCV criterion. %a Craven-Wahba type estimate.
Therefore, we replace Eq.~(A10) with a modified loss estimator based on
the Rice  criterion:
$$ %\begin{equation}\label{A13}
 C_R \equiv
 \frac{||\ybf -\Xbf  \hat{\albf}_{\lambda} ||^2_{\Dbf}}
{N- 2 {\rm trace}[\Kbf \Gbf] }  \ .\eqno (A13)
$$
$C_R$ has also been normalized to the standard error per data point.
$C_R$ differs from Eq.~(A12) by terms of $O({\rm trace}[\Kbf \Gbf] / N)$. 
%i.e.~$C_R$ also differs from Eq.~(A6) by a factor of $\frac{1}{\sigma^2N}$.
%$$ C_R \equiv
% \frac{|{\bf y}-{\bf X} \hat{\alpha}_{\lambda} |^2}{\sigma^2N}  -2
%+ \ \frac{N}{N- 2 {\rm trace}[\Cbf \Gbf] }  \ .\eqno (A7)$$
% In Ref.~12, a different empirical estimate of EASE is used to 
% determine the smoothing parameters. The EASE estimate in Ref.~12 does not 
% include the model error in the directions perpendicular to the 
% column space of $\Xbf$ and therefore is not appropriate for model
% selection.
In our analysis, we minimize Eq.~(A13) with respect to both the choice of
control variables in the additive model and the smoothing parameters in a given
model. In Ref.~24,
 a somewhat different autocorrelation correction is  derived.

An older statistic is $\chi^2 \equiv
 \frac{||\ybf -\Xbf  \hat{\albf}_{\lambda} ||^2}
{N-  {\rm trace}[\Kbf \Gbf] }$,
which corresponds to the mean square error per degree of freedom. 
The $\chi^2$ statistic is useful in optimizing the fit
to existing data while the Rice criterion and generalized  crossvalidation
are useful in minimizing the predictive error for new data.
The factor of two in the denominator of $C_R$ results in smoother models 
and fewer variables in the model.

\

\cl{ACKNOWLEDGMENTS}

\medskip
G.~Cordey's support and encouragement are gratefully acknowledged.
We thank %the JET team; in particular 
C.~Gowers, P.~Nielsen, K.~Thomsen, and D.~Muir.
KI's work was supported by the U.S. Department of Energy Grants 
No.DE-FG02-92ER54157.
KSR's work was supported by the U.S. Department of Energy Grants 
DE-FG02-86ER-53223 and 91ER54131.

\np
\begin{center}
{\bf REFERENCES}
\end{center}

\begin{enumerate}
%``Status of global confinement studies."
%.~M.~Kaye, M.~G.~Bell, et al.,~{Physics of Fluids B} {\bf 2} 2926 (1990).
\item K.~S.~Riedel, S.~M.~Kaye, { Nuclear Fusion}
{\bf 30}  731 (1990).
\item  K.~S.~Riedel,
%``Tokamak to tokamak variation and colinearity in scaling laws,''
Nuclear Fusion {\bf 30}  755 (1990).
%\item Goldston, R.~J., Bateman, G.,  Bell, M.~G. et al. 
%{\it``Performance Projections for B.P.X.",}
%Bull. of Am. Phys. Soc. {\bf 35}, No. 9,  poster 1P1, (1990) 1920.
\item  K.~S.~Riedel,
%``Advanced Statistical Techniques for Tokamak Data Analysis",
Comments in Plasma Physics and Controlled Fusion
{\bf 12}  (1989) 279.
\item P.~Yushmanov, T.~Takizuka,~~K.~S.~Riedel,~~O.~J.~Kardaun, J.~G.~Cordey,
S.~Kaye and D.~Post, 
Nuclear Fusion {\bf 30} (1990) 1999.
\item  K.~S.~Riedel,
%`` On dimensionally correct power law scaling
%expressions for L-mode confinement,"
{ Nuclear Fusion}, {\bf 31}   927 (1991).
\item  J.~P.~Christiansen, J.~G.~Cordey,  O.~J.~Kardaun,
and K.~Thomsen, Nuclear Fusion, {\bf 31}   2117 (1991).
\item  J.~P.~Christiansen,  J.~G.~Cordey, K.~Thomsen, A.~Tanga 
and the JET team,  J.~C.~DeBoo, D.~P.~Schissel, T.~S.~Taylor,
 and the DIII-D team,
O.~J.~Kardaun,  F.~Wagner, F.~Ryter and the ASDEX team, S.M.~Kaye and
the PDX and PBX-M teams, Y.~Miura and the JFT-2M group
{Nuclear Fusion}, {\bf 32}  (1992) 291.
%``A Global Energy Confinement H-mode Database for I.T.E.R.",
%submitted to {Nuclear Fusion}.

\item  M.~H.~Redi,  W.~M.~Tang, P.~C.~Efthimion,
D.~R.~Mikkelsen, G.~L.~Schmidt,
{Nuclear Fusion} {\bf 27}  2001 (1987).
\item   W.~M.~Tang,
{ Nuclear Fusion}, {\bf 26}   1605 (1986).

\item{ P.~J.~ McCarthy, K.~S.~Riedel,  O.~J.~Kardaun,
H.~Murmann, K.~Lackner, 
Nuclear Fusion {\bf 31} (1991) 1595,
also {\it Scalings and Plasma Profile Parameterisation of ASDEX High Density
Ohmic Discharges},
Max-Planck-Institut f\"ur Plasmaphysik Report No. 5/34.}

\item O.~J.~Kardaun,  K.~S.~Riedel, P.~J.~McCarthy and K.~Lackner, 
 Max-Planck-Institut f\"ur Plasma Physik Report No. 5/35,
(1990).

\item  K.~S.~Riedel and K.~Imre, 
Comm.~in Statistics {\bf 22},   1795 (1993).

\item P.~H.~Rebut, R.~J.~Bickerton, B.~E. Keen, Nuclear Fusion {\bf  25}
1011, (1985).

\item{
P.~Craven and  G.~Wahba %~(1979).
%Smoothing noisy data with spline functions: estimating the correct
%  degree of smoothing by the method of generalized cross-validation.
{ Numer.~Math.}~{\bf 31},  (1979) 377.} %-403.}

\item{W.~Hardle, P.~ Hall, and S.~Marron, 
%{How far are automatically chosen smoothing parameters from their optimum?}
{ J.~Amer.~Statist. Assoc.} {\bf 83} (1988) 86.}%-95

\item H.~Salzmann, J.~Bundgaard, A.~Gadd, C.~Gowers,  P.~Nielsen,
Rev.~Sci.~Instrum. {\bf 59}, (1988) 1451.
%{\it The LIDAR Thomson scattering diagnostic on JET.}
%JET Report JET-R (89)-07 (1989). 

\item{V.~D.~Arunsalam,  N.~L.~Bretz,  P.~C.~Efthimion, R.~J.~Goldston, B.~Grek,
D.~W.~Johnson, M.~Murakami, K.~M.~McGuire, D.~A.~Rasmussen, F.~J.~Stauffer
and J.~B.~Wilgen
Nuclear Fusion {\bf 30} (1990) 2111.}

\item{
 C.~K.~Phillips,  W.~Houlberg, D.~Hwang,  
S.~Attenberger, J.~Tolliver, L.~Hively, 
%"Predictive Transport Modeling of ICRF-Heated Tokamaks,"
Plasma Phys.~Control.~Fusion {\bf 35} (1993) 301. %-317.
}

\item{J.~P.~Christiansen,   J.~D.~Callen,  J.~G.~Cordey, K.~Thomsen,  %et al.
 Nuclear Fusion {\bf 28}  (1988) 817.}

\item{ 
%J.~A.~Murphy,  S.~D.~Scott,  H~.H.~Towner, {\it The SNAP user's Guide}
%{Princeton Plasma PhysicsLaboratory Technical Manuscript \# 393.}}(1992).
H.~H.~Towner, R.~J.~Goldston, G.~W.~Hammett, J.~A.~Murphy,
C.~K.~Phillips, S.~D.~Scott, M.~C.~Zarnstorff, D.~Smithe,
Reviews of Sci.~Inst.  {\bf 63},4753-6, (1992).}
%ruments)%  year={1992},  volume={63},  number={10}}

\item{ O.~Gruber,
Nuclear Fusion {\bf 22},  (1982) 1349.}

\item D.~J.~Grove, D.~M.~Meade,  Nuclear Fusion {\bf  25}
1167, (1985).

\item D.~Post et al.  {\em I.T.E.R.~Physics Basis.} 
{I.A.E.A. Publishing}, Vienna, (1991).

\item{ N.~S.~Altman,
%``{Kernel smoothing with correlated errors},'' 
{J.~Amer.~Stat.~Assoc.} {vol.~85}, (1990) pp.~749. %-759, (1990)}
}

%``Scaling expressions for ITER,''
%ITER Report IL-Ph-4-9-7,  Nucl. Fusion, {\bf 30}, p.1999, (1990).

%\item Kadomtsev, B.B., Sov. J. Plasma Phys., 295, (1975). 

\end{enumerate}

\includepdf[pages=-,pagecommand={}]{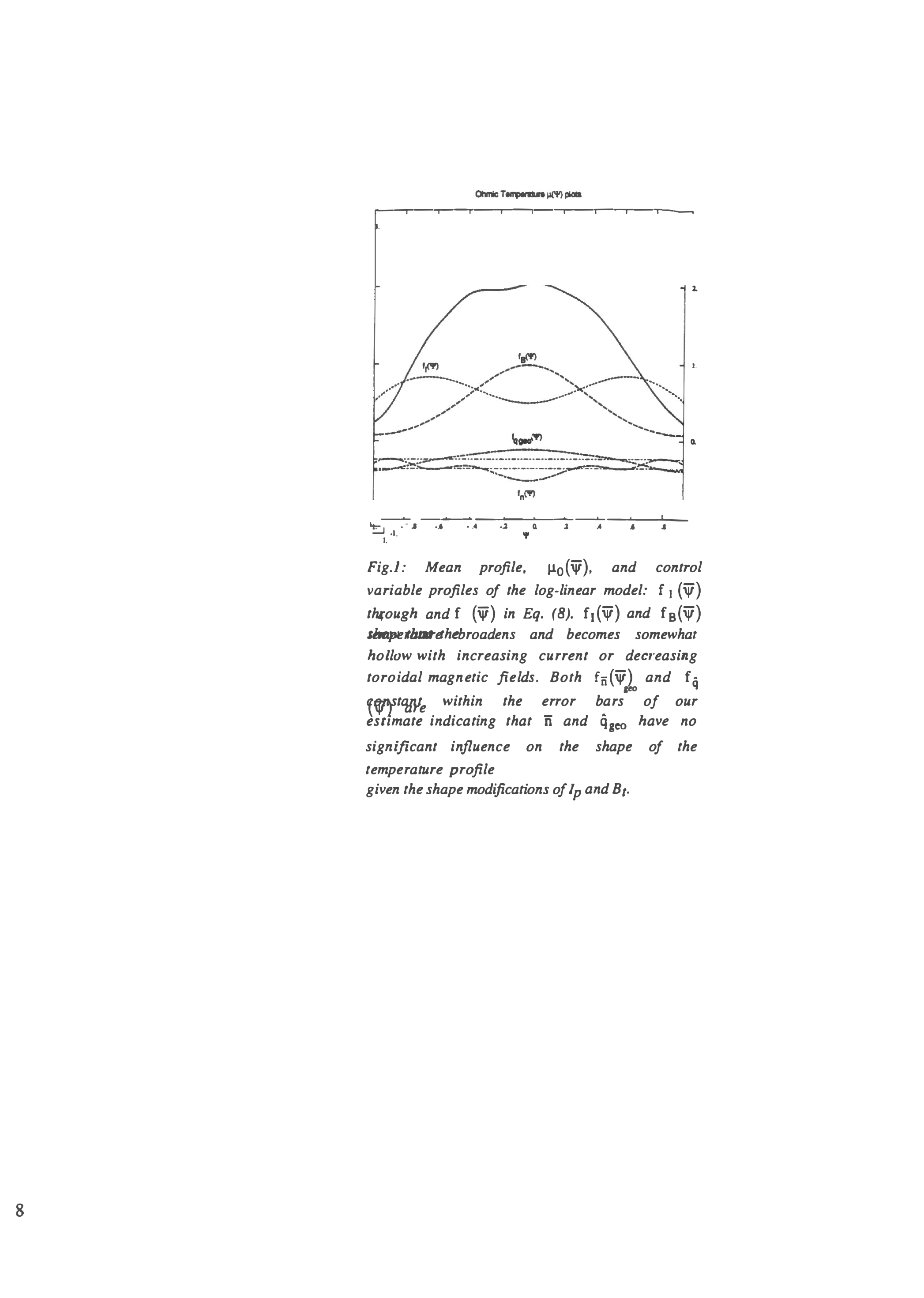}

\cl{\bf TABLES}

%\cl{Database Summary}

\

\begin{center}
\begin{tabular}{|r|c|c|c|c|}
\hline
         Var & mean &min & max &std dev \\ 
\hline
$\bar{n}$    & 2.29  &  1.32    & 3.90   & 0.75     \\
$q_{95}$     & 5.48  &  2.88    & 12.6   & 2.86     \\
$I_p$        & 2.80  &  0.97    & 5.25   & 1.13   \\
$B_t$        & 2.76  &  1.30    & 3.22   & 0.46     \\
$\kappa$     & 1.44  &  1.30    & 1.75   & 0.122    \\
$a$          & 1.16  &  1.05    & 1.19   & 0.040     \\   
$R$          & 2.92  &  2.83    & 3.01   & 0.047     \\
Volt         & -.35  & -1.12    & .914   & 0.66     \\
$Z_{eff,1}$  & 1.88  &  1.07    & 3.10   & 0.55     \\
$Z_{eff,2}$  & 2.10  &  1.20    & 3.35   & 0.60      \\
\hline
\end{tabular}
\end{center}

\

Table 1: Database Summary:
Average, minimum,  maximum and 
standard deviation of each of the engineering variables.

 \

\np

\begin{center}
{\em Sequential Selection Using Rice Criterion}

\

\begin{tabular}{|rccccc|}
\hline
Vars in model & 1 Var& 2 Var& 3 Var & 4 Var & 5 Var \\
\hline
$\ln[\bar{n}]$& 4.64 & 1.73 &\underline{1.12} & seed & seed \\
$\ln[q_{95}]$ & 3.04 & 1.78 & 1.36 &\underline{ .885} & seed \\
$\ln[I_p]$&\underline{\it 1.93}& seed & seed & seed & seed \\
 Volt         & 4.63 & 1.92 & 1.53 & 1.10 & .861 \\
$\ln[B_t]$    & 3.96 &\underline{1.58} & seed & seed & seed \\
$\ln[\kappa]$ & 4.30 & 1.94 & 1.56 & 1.10 & .875 \\
$\ell_i$      & 4.21 & 1.63 & 1.48 & 1.05 & .875 \\
 $a$          & 4.60 & 1.91 & 1.58 & 1.11 & .865 \\
 $R$          & 4.47 & 1.88 & 1.58 & 1.06 & .869 \\
$Z_{eff,1}$   & 4.01 & 1.91 & 1.55 & 1.06 & .872 \\
$Z_{eff,2}$   & 4.37 & 1.79 & 1.57 & 1.11 & .850 \\
%12-goeq   & 4.66 & 1.66 & 1.36 & .886 & blow  seed
Time          & 4.58 & 1.87 & 1.52 & .923 & .793 \\
%13-I/a    & 1.96 & 1.91 & 1.58 & 1.11 & .868 & .8687
%14-Ba/I   & 3.06 & 1.58 & 1.57 & 1.11 & .867 & .8681 
\hline
\end{tabular}

\end{center}

Table 2: Rice criterion as a function of the variables in the model.
``Seed variables'' are included in each run in that  column. We then add the
variable that reduces  the criterion the most.

%\end{document}

\np

\cl{Fitted Functions for Eq.~(\ref{E8})}

\

\begin{center}
\begin{tabular}{|rccccc|}
\hline
$\psb$ & $f_0(\psb$) & $f_I(\psb)$ &$f_B(\psb$) & $f_n(\psb)$ & 
$f_{\qh}(\psb$) \\
\hline
 -1.0  & 0.2376  & 0.5057  & 0.0776  & -0.3013  & -0.3879 \\
 -0.9  & 0.4267  & 0.6679  & 0.0900  & -0.2332  & -0.3755 \\
 -0.8  & 0.7972  & 0.7728  & 0.1320  & -0.2710  & -0.3370 \\
 -0.7  & 1.1884  & 0.8231  & 0.2037  & -0.3479  & -0.2902 \\
 -0.6  & 1.4813  & 0.8236  & 0.3046  & -0.3746  & -0.2484 \\
 -0.5  & 1.7071  & 0.7827  & 0.4317  & -0.3449  & -0.2161 \\
 -0.4  & 1.8869  & 0.7131  & 0.5771  & -0.3294  & -0.1891 \\
 -0.3  & 1.9517  & 0.6316  & 0.7261  & -0.3658  & -0.1631 \\
 -0.2  & 1.9528  & 0.5561  & 0.8577  & -0.4384  & -0.1397 \\
 -0.1  & 1.9785  & 0.5029  & 0.9491  & -0.5021  & -0.1236 \\
 0.0  & 2.0271  & 0.4838  & 0.9820  & -0.5261  & -0.1179 \\
 0.1  & 2.0257  & 0.5029  & 0.9491  & -0.5021  & -0.1236 \\
 0.2  & 1.9588  & 0.5561  & 0.8577  & -0.4384  & -0.1397 \\
 0.3  & 1.8611  & 0.6316  & 0.7261  & -0.3658  & -0.1631 \\
 0.4  & 1.7285  & 0.7131  & 0.5771  & -0.3294  & -0.1891 \\
 0.5  & 1.5236  & 0.7827  & 0.4317  & -0.3449  & -0.2161 \\
 0.6  & 1.2578  & 0.8236  & 0.3046  & -0.3746  & -0.2484 \\
 0.7  & 0.9711  & 0.8231  & 0.2037  & -0.3479  & -0.2902 \\
 0.8  & 0.6821  & 0.7728  & 0.1320  & -0.2710  & -0.3370 \\
 0.9  & 0.4173  & 0.6679  & 0.0900  & -0.2332  & -0.3755 \\
 1.0  & 0.2243  & 0.5057  & 0.0776  & -0.3013  & -0.3879 \\
\hline
\end{tabular}
\end{center}
%------------------------------------------
%nb= exp(0.7753), qb= exp(1.5123), Ib= exp(0.9368), Bb= exp(0.9968)

Table 3: Evaluation of the radial spline functions in Eq.~(\ref{E8}) 
for the JET data.

\np

\cl{Fitted Functions for Eq.~(\ref{E8Q})}

\

\begin{center}
\begin{tabular}{|rccccc|}
\hline
$\psb$ & $f_0(\psb$) & $f_q(\psb)$ &$f_I(\psb$) & $f_B(\psb)$ & 
$f_n(\psb$) \\
\hline
 -1.0  & 0.2326  & -0.3729  & 0.6868  & 0.4900  & -0.3652 \\
 -0.9  & 0.4279  & -0.3364  & 0.6868  & 0.4900  & -0.3652 \\
 -0.8  & 0.8015  & -0.2822  & 0.6868  & 0.4900  & -0.3652 \\
 -0.7  & 1.1914  & -0.2298  & 0.6868  & 0.4900  & -0.3652 \\
 -0.6  & 1.4827  & -0.1951  & 0.6868  & 0.4900  & -0.3652 \\
 -0.5  & 1.7092  & -0.1836  & 0.6868  & 0.4900  & -0.3652 \\
 -0.4  & 1.8890  & -0.1885  & 0.6868  & 0.4900  & -0.3652 \\
 -0.3  & 1.9514  & -0.1995  & 0.6868  & 0.4900  & -0.3652 \\
 -0.2  & 1.9505  & -0.2105  & 0.6868  & 0.4900  & -0.3652 \\
 -0.1  & 1.9750  & -0.2186  & 0.6868  & 0.4900  & -0.3652 \\
 0.0  & 2.0194  & -0.2215  & 0.6868  & 0.4900  & -0.3652 \\
 0.1  & 2.0154  & -0.2186  & 0.6868  & 0.4900  & -0.3652\\
 0.2  & 1.9511  & -0.2105  & 0.6868  & 0.4900  & -0.3652\\
 0.3  & 1.8579  & -0.1995  & 0.6868  & 0.4900  & -0.3652\\
 0.4  & 1.7298  & -0.1885  & 0.6868  & 0.4900  & -0.3652\\
 0.5  & 1.5275  & -0.1836  & 0.6868  & 0.4900  & -0.3652\\
 0.6  & 1.2613  & -0.1951  & 0.6868  & 0.4900  & -0.3652\\
 0.7  & 0.9729  & -0.2298  & 0.6868  & 0.4900  & -0.3652\\
 0.8  & 0.6841  & -0.2822  & 0.6868  & 0.4900  & -0.3652\\
 0.9  & 0.4194  & -0.3364  & 0.6868  & 0.4900  & -0.3652\\
 1.0  & 0.2208  & -0.3729  & 0.6868  & 0.4900  & -0.3652 \\
\hline
\end{tabular}
\end{center}
%------------------------------------------
%nb= exp(0.7753), qb= exp(1.5123), Ib= exp(0.9368), Bb= exp(0.9968)

Table 4: Evaluation of the radial spline functions in Eq.~(\ref{E8Q}) 
for the JET data.

\end{document}